\documentclass[12pt]{iopart}
\usepackage{amssymb}
\usepackage{epsf}

 \newcommand{\bm}[1]{\mbox{\boldmath $#1$}}

 \newcommand{\zr}[1]{\mbox{\hspace*{#1em}}}
 \newcommand{\nqv}{\equiv\!\!\!\!\!/\;}

 \newcommand{\EE}{\mathbb E}
 \newcommand{\NN}{\mathbb N}
 \newcommand{\RR}{\mathbb R}
 
 \newcommand{\ZZ}{\mathbb Z}

 \newcommand{\QQ}{\mathbb Q}

 \newcommand{\be}{\begin{equation}}
 \newcommand{\ee}{\end{equation}}

\begin{document}

\title[$p$-adic quasicrystals]{Limit-(quasi-)periodic point sets as 
quasicrystals with $p$-adic internal spaces}

\author{Michael Baake\footnote{Heisenberg Fellow}$^1$, Robert V. Moody$^2$
\\ and 
Mar\-tin Schlott\-mann$^2$}

\address{{}$^1$ Institut f\"ur Theoretische Physik, Universit\"at T\"ubingen,\\
 Auf der Morgenstelle 14, D-72076 T\"ubingen, Germany}

\address{{}$^2$ Department of Mathematical Sciences, University of Alberta,\\ 
 Edmonton, Alberta T6G 2G1, Canada}

\vspace{1.5cm}
\hspace*{16mm}
Dedicated to Peter Kramer on the occasion of his 65th birthday
\vspace{1cm}

\begin{abstract}
Model sets (or cut and project sets) provide a familiar and commonly used  
method of constructing and studying nonperiodic point sets. Here we extend this 
method to situations where the internal spaces are no longer 
Euclidean, but instead spaces with $p$-adic topologies or even
with mixed Euclidean/$p$-adic topologies.
We show that a number of well known tilings precisely fit this form, including
the chair tiling and the Robinson square tilings. Thus the scope of the cut and
project formalism is considerably larger than is usually supposed. Applying
the powerful consequences of model sets we derive
the diffractive nature of these tilings.

\end{abstract}

 


\maketitle

\section{Introduction}

The cut and project method of constructing nonperiodic point sets,
as developed by Peter Kramer and others in the early eigthies 
\cite{Kramer,KN,Katz,deBr,McM}, is one of the basic tools in the mathematical 
study of quasicrystals and aperiodic order. The intuition behind their
use is that quasiperiodic point sets, such as those arising in 
many nonperiodic tilings and also in the diffraction patterns
of physical quasicrystals, may be viewed as arising from the projection 
of lattices in some higher dimensional spaces. Thus the physical
space is complemented by an internal space (possibly of some other
dimension), a lattice is located in the combined physical-internal 
space pair, and the projection maps are used to create a cut and 
project scheme.

The same type of mathematical structure had also arisen (before
the recent excitement about quasicrystals, and in a very different
context) in the work of Yves Meyer
\cite{YM1} in which the formalism is expressed entirely in terms
of locally compact Abelian groups. In \cite{RVM,Martin2}
these ideas were taken up and extended in the context of aperiodic
order, with the result that a considerable amount of the mathematical
theory underlying these cut and project sets (or model sets) can now
be seen to hold in great generality. Up to now, however, no attempt
has actually been made to see to what extent existing aperiodic structures
might be explained in terms of these more general types of spaces. In this 
paper we address this question, showing that a number of familiar
tilings and substitution systems, so far not contained under the 
aegis of the cut and project formalism, are in fact based on internal spaces
with {\em non-Euclidean} topologies, namely $p$-adic topologies
or mixed Euclidean/$p$-adic topologies.

There is a variety of discrete structures known that display a pure
 point diffraction spectrum, i.e., the Fourier transform of their
 autocorrelation (which is a positive measure) is pure point,
 compare \cite{bert,boris}.
Among the known examples are model sets, but also certain
 inflation-generated point sets and tilings, e.g.\ the chair or the
 sphinx tiling \cite{boris}.
They are limit-periodic structures with a countably, but not finitely,
 generated Fourier module -- so, they cannot be described in the
 ``conventional'' cut and project setup where the internal space is
 Euclidean.

It is the aim of this contribution to start to develop a proper
 generalization of the projection method, in the spirit of Meyer,
 to include such limit-periodic and even limit-quasiperiodic sets.
Here, we explain, in an illustrative fashion, how this works.
A detailed approach to model sets over arbitrary internal 
groups, and especially to aspects of diffraction, will appear
in \cite{forthcoming}.

Let us recall the notion of a cut and project scheme. By definition,
this consists of a collection of spaces and mappings: 

\be \label{cutandproject}
  \begin{array}{ccccc}
   \RR^d & \stackrel{\pi^{}_1}{\longleftarrow} & \RR^d \times G &
           \stackrel{\pi^{}_2}{\longrightarrow} & G  \\
    & & \cup & & \\ & & L & & \end{array}
\ee
where $\RR^d$ is a real Euclidean space and $G$ is some
locally compact Abelian group, $\pi^{}_1$ and 
$\pi^{}_2$
are the projection maps onto them, and $L \subset 
\RR^d \times G$
is a lattice, i.e., a discrete subgroup such that the quotient group
 $(\RR^d\times G)/L$ is compact. We assume that $\pi^{}_1|^{}_L$ is
 injective and
that $\pi^{}_2(L)$ is dense in $G$. We call $\RR^d$
(resp.\ $G$) the physical (resp.\ internal) space. 

In this definition, we have already oriented the situation 
to physical applications by assuming that the physical space
is indeed a real Euclidean space. On the other hand, allowing
$G$ to be an arbitrary locally compact Abelian group is precisely
the point at which we are going beyond the usual situation of an internal space
that is also Euclidean.

Given any subset $\Omega \subset G$, we define a corresponding 
set $\Lambda(\Omega) \subset \RR^d$ by
\be
  \Lambda (\Omega)\; = \; \{\pi_1(x)  \;|\; x \in L , \, \pi_2(x) \in \Omega \} \,.
\ee 
We call such a set $\Lambda$ a {\em model set} 
(or {\em cut and project set}) if the following condition is fulfilled,
\begin{itemize}
\item[\bf W1] $\Omega \;   =  \; \overline{\mbox{int}(\Omega)}
                      \; \neq \; \emptyset\;$
 is compact.
\end{itemize}

Furthermore, we are mainly interested in the situation that the boundary
 of $\Omega$ does not contain any points of $\pi_2(L)$. If this is the
 case, we call $\Lambda(\Omega)$ {\em regular}. The importance of regular
 model sets is that they are necessarily repetitive \cite{Martin2}.
Note that, if $\Omega$ fulfils {\bf W1}, 
 $\partial\Omega$ (which equals $\Omega\backslash\mbox{int}(\Omega)$) 
 is nowhere dense and hence a meager set.
Then, it follows from the Baire category theorem
 that no countable union of translates of $\partial\Omega$ can 
 cover $G$ which is a Baire space.
In particular, it is always possible to choose a shift
 $c\in G$ such that the boundary of $c+\Omega$ satisfies the additional
 regularity condition. 

In the sequel, showing that a certain set is a model set actually means, more
 precisely, to show that there exist $G$, $L$ and $\Omega$ subject
 to the above conditions such that $\Lambda(\Omega)$ is regular and locally
 isomorphic to the given set (for terminology, see \cite{BS,Martin2} and 
 references therein). This way, our results are valid for entire LI-classes,
 even if, for simplicity, we only talk of single tilings.

\section{$p$-adic topologies and inverse limits of finite groups}

Let $p$ be a prime number in the integers $\ZZ$. Using $p$ we can 
define a metric on the rational numbers $\QQ\/$, and
by restriction on $\ZZ$, in the following way. For each $a \in \ZZ$,
we define its $p$-value, $\nu_p(a)$, as
the largest exponent $k$ for which $p^k$ divides $a$ 
(with $\nu_p(0):= \infty$). This function
is extended to the $p$-adic valuation $\nu_p: \QQ \longrightarrow \ZZ$ by 
$\nu_p(a/b) := \nu_p(a) - \nu_p(b)$ for all rational numbers
$a/b$. We now define the ``distance'' between two rational numbers $x,y$
as $ d(x,y) = p^{-\nu_p(y-x)}$ .

It is not hard to see that this does define a metric on $\QQ$, in which 
closeness to $0$ is equivalent to high divisibility by the prime
$p$. The completion of the rationals under this topology
is the field of $p$-adic numbers $\widehat{\QQ_p}$ and the 
completion of $\ZZ$ is the subring of $p$-adic integers, $\widehat{\ZZ_p}$.
Any $p$-adic integer has a unique expansion (as a convergent series)
in the form $\sum_{n=0}^\infty a_np^n$
where the $a_p$ are integers in the range $0 \le a_p < p$.
The topologies defined by such metrics 
have rather counter-intuitive properties. For example,
for each non-negative integer $k$, the set $p^k\cdot\widehat{\ZZ_p}$, 
the set of elements of $\widehat{\ZZ_p}$ divisible by $p^k$, is the ball 
of radius $p^{-k}$ and is clopen, i.e.\ both open and closed. $\widehat{\ZZ_p}$, 
seen as a topological space, is both compact and totally disconnected.
In particular, $\widehat{\QQ_p}$ and $\widehat{\ZZ_p}$
are locally compact Abelian groups under addition.
Thus, we can use $\widehat{\ZZ_p}$ to construct interesting cut and project 
schemes for $\RR^d$ simply by taking $G := \widehat{\ZZ_p}$ and $L = \ZZ^d$  
embedded diagonally into $\RR^d \times  \widehat{\ZZ_p}$.
For more on $p$-adic numbers and other totally disconnected
groups, the reader may consult \cite{Neukirch,Serre,HR,Bourbaki}.

There is another description of  $\widehat{\ZZ_p}$ which is more 
revealing of its appearance in the context of self-similarity and generalizes
what we have just done. Let 
\be
  F_1 \, \leftarrow \, F_2  \, \leftarrow \, F_3 \leftarrow \dots 
\ee
be an inverse system of finite Abelian groups, i.e.\ each $F_i$ is a finite
Abelian group (with discrete toplogy) and the arrows represent surjective 
group homomorphisms. Define the set of compatible sequences
\be
\overleftarrow{F} \; := \;
 \{ \tilde{x} = (x_1, x_2,  \dots ) \mid 
     x_i \in F_i , \, x_i \, 
         \leftarrow\mbox{\raisebox{0.13em}{$\hspace*{-0.43em} \scriptscriptstyle |$}} 
                   \, \,   x_{i+1}, \, i \in \NN \} \, .
\ee
$\overleftarrow{F}$ is given the structure of a group by 
component-wise addition. It is
structured as a topological group by the induced topology from the 
product $\prod_{i=1}^{\infty} F_i$. Equivalently, the subgroups
\be
\overleftarrow{F}_{\!\!n} \; = \; \{\tilde{x} \in \overleftarrow{F}
  \; \mid \; x_1 = x_2 = \dots = x_{n-1} = 0 \}
\ee
form a subbase  of open neighbourhoods of $0$ in $\overleftarrow{F}$.
Since $[F:F_n]$
is finite, the subgroups $\overleftarrow{F}_n$ are also closed. 
With this topology,
$\overleftarrow{F}$ is a compact totally disconnected Abelian
group (so, in particular, a locally compact Abelian group). 
Groups of this type are called profinite groups.

As an example, for each prime number $p$, we can construct $\widehat{\ZZ_p}$
by the inverse system
\be 
     \overleftarrow{\,\ZZ}_{p} \; : \quad
               \ZZ/p\ZZ \, \leftarrow \, \ZZ/p^2 \ZZ
                        \, \leftarrow \, \ZZ/p^3 \ZZ 
                        \, \leftarrow \, \dots 
\ee
 
The relevance to the work here is this: if $\theta : L \rightarrow L$
is an injective homomorphism that is a self-similarity of $\Lambda$,
then there is a clear distinction between the case that $\theta(L) = L$
($\theta$ is a ``unit'') and the case that 
$\theta(L) \subset L$, but $\theta(L)\neq L$. 
In the latter case, $[L:\theta(L)]$ is finite and we have the
inverse system 
\be \label{profinite}
\overleftarrow{L}(\theta) \; : \quad
      L/\theta(L) \, \leftarrow \, L/\theta^2(L) 
                  \, \leftarrow \, L/\theta^3(L) 
                  \, \leftarrow \, \dots  
\ee
The compact group $\overleftarrow{L}$ is invariant under the action of $\theta$.
Note that it contains a canonical copy of $L$ itself via the mapping
\be
 x \; \mapsto \; 
   ( ( x \; \mbox{ mod} \, \theta(L)   ) \, 
  \leftarrow\mbox{\raisebox{0.13em}{$\hspace*{-0.43em} \scriptscriptstyle |$}} \;
     ( x \; \mbox{ mod} \, \theta^2(L) ) \, 
  \leftarrow\mbox{\raisebox{0.13em}{$\hspace*{-0.43em} \scriptscriptstyle |$}} \;
     ( x \; \mbox{ mod} \, \theta^3(L) ) \, 
  \leftarrow\mbox{\raisebox{0.13em}{$\hspace*{-0.43em} \scriptscriptstyle |$}} \; 
     \dots  )  \, .
\ee
Again, we obtain a cut and project scheme via the diagonal embedding of $L$
in $\RR^m \times \overleftarrow{L}$. 
Let us now turn to some applications.

\section{A limit-periodic substitution system}

Consider the primitive three-letter substitution system 
\be  \label{substRule} 
 a \rightarrow ab \qquad b \rightarrow abc
\qquad c \rightarrow abcc
\ee
The standard analysis of the corresponding substitution matrix \cite{queff}
 shows that a proper geometric realization demands length ratios
 $\ell(a):\ell(b):\ell(c)=1:2:3$, while all three letters finally occur with 
 equal frequency $1/3$.

To obtain a bidirectional infinite sequence that is a fixed point, 
 we may start with the pair $c|a$ and keep on applying the substitution rule:
\be
\dots \, ab abc abcc abcc | ab abc ab abc abcc ab abc ab abc abcc \, \dots 
\ee
We can imagine this as labelling the tiles of a tiling of $\RR$
in which the tiles are of lengths $1,2,3$ respectively.
If we identify each tile with its right-hand end point,
starting with a tile of type $a$ (length $1$) at the origin,
then we end up with a sequence of numbers
\be
  \mbox{ \small{\dots 
-26,-24,-21,-18,-17,-15,-12,-9,-8,-6,-3,0,1,3,4,6,9,10,12,13,15,18,19 
  \dots }}
\ee

The main property of this sequence is its invariance under
the transformation $x \mapsto 3x$. This self-similarity with 
a {\em rational} scaling factor is the signal that there may
be a $p$-adic interpretation. At the same time, as 3 is not a unit,
this sequence is a candidate for a so-called limit-periodic point
set, compare \cite{GK}.
In fact, this sequence can be given a $3$-adic interpretation. 
The coordinates of the tiles of the three types (resp.\ their right hand 
endpoints) can be explicitly given as follows:
\begin{itemize}
\item  type $a$ : \quad $\bigcup_{k=2}^\infty (1 + 3 + \dots +
3^{k-2}) \, + \,3^k\ZZ$
\item  type $b$ : \quad $\bigcup_{k=2}^\infty (2 + 1 + 3 + \dots +
3^{k-2})\,  + \, 3^k\ZZ$
\item type $c$ : \quad $3^2 \ZZ  \; \cup \; 
\left(\bigcup_{k=3}^\infty 
(- 3 - \dots - 3^{k-2}) \, + \, 3^k\ZZ \right)$.
\end{itemize}
It is easy to see that these sets are invariant under the process
of formation of the tiles (i.e., rule (\ref{substRule})). 
Furthermore, their densities in the 
lattice of integers $\ZZ$ are easily computed to be each equal
to $1/6$, thus accounting for the entire tiling
 ($(1 +2 +3)\cdot\frac{1}{6} = 1$).

Interpreting these sets $p$-adically, we see that they are dense subsets
of the unions of open balls formed by replacing $3^k\ZZ$ by
$3^k\widehat{\ZZ_3}$ in each of the sets above.
In this way, we obtain three ``windows'',
each with compact closure. Furthermore, it is not hard
to see that each of them has a boundary with just finitely
many points.

Using the lattice $L:=\{ (n,n) | n\in \ZZ \}\subseteq \RR\times \widehat{\ZZ_3}$
we obtain a cut and project scheme and conclude that the point sets corresponding
to each of the three tile types is a model set.
As a consequence, the sequence is an example with pure point
 diffraction spectrum.
This observation is consistent with the following application
 of Dekking's criterion \cite{dekking} to a locally equivalent sequence
 \cite{BS}, which would share pure pointness with our above example due to 
 the invariance of this spectral property under mutual local derivability.
Indeed, replacing the configuration $ab$ by a new tile $A$ of length 3 one
 obtains a sequence invariant under the substitution rule
 $A\rightarrow AAc$, $c\rightarrow Acc$ which  has a pure point
 diffraction spectrum because the new substitution rule is of constant length
 and exhibits a so-called coincidence \cite{dekking}.

\section{The chair tiling}

The 2D chair tiling is defined by the substitution rule in Figure 1a.
This has recently been shown to display a pure point diffraction spectrum \cite{boris}.
Instead of working with the chair tiling directly, we introduce
 a convenient modification by substituting each ``chair'' by
 three decorated squares as in Figure 1b.
Since this transformation is local in the sense of \cite{BS} and
 can be locally inverted by the rule given in Figure 1c, it follows
that showing that the chair tiling is a cut and project tiling is equivalent
 to showing that the modified tiling is.
With the help of the transformation rules, one sees immediately
 that the modified tiling fulfils the substitution rule in
 Figure 1d.
 
\begin{figure}
\vspace*{5mm}
\centerline{\epsfysize=65mm \epsfbox{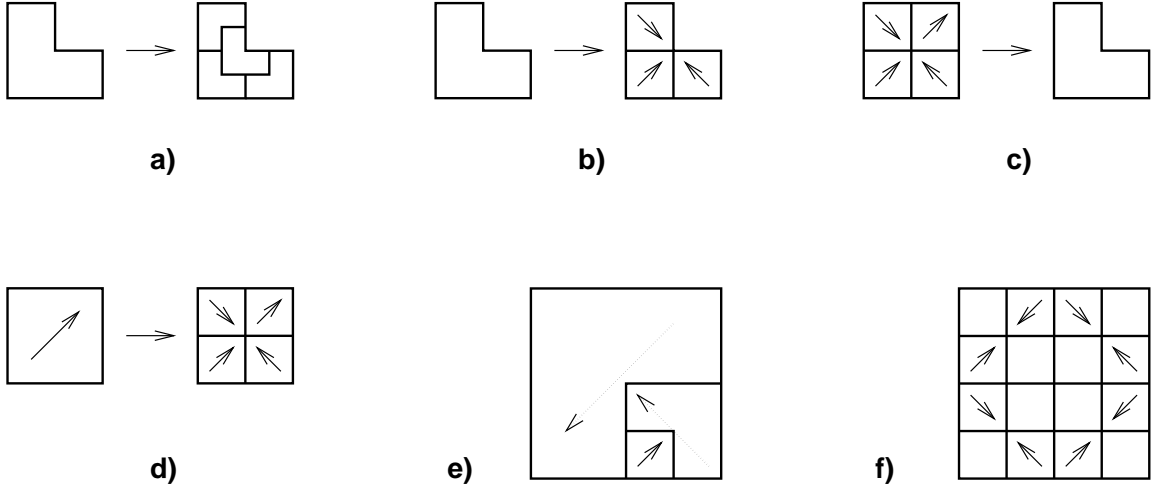}}
\caption{Geometric realization of the chair tiling.}\label{fig1}
\end{figure}

A particular tiling ${\cal T}_0$ in the LI class under consideration
 can be defined in the following way.
Starting with a decorated square $S_0$ of side length 1 centered at the origin
 of a fixed coordinate system where the arrow points towards the upper
 right corner $(1/2,1/2)$, one performs successively the following
 two steps: 
\begin{enumerate}
\item perform the affine transformation
\be
T: \quad x \; \mapsto \; Tx:= 2Rx+\frac{1}{2} (\vec{\bm{e}}_1+\vec{\bm{e}}_2),
\ee
 where $R$ denotes rotation by $\pi/2$ and $\vec{\bm{e}}_i$ are the
 canonical unit vectors;
\item apply the (appropriately rotated) substitution rule of Figure 1d.
\end{enumerate}
This way, larger and larger portions of a unique member of the LI class
 are obtained (see Figure 1e).

As a consequence of the construction, the centers of the decorated
 squares form the lattice $\ZZ^2$.
Let $P_k$ be the subset of the centers of squares oriented as $R^k S_0$,
 $k=0,1,2,3$, respectively.
Clearly, the tiling is completely determined by these subsets.

We will show that each $P_k$ is a model set with internal group 
 $G=\widehat{\ZZ_2} \times \widehat{\ZZ_2}$.
As in our previous one-dimensional example, this will be done by writing
 $P_k$ as a union of cosets of certain sublattices of $\ZZ^2$.
Because of the partial symmetry of the decoration of the
 second substitution step (see Figure 1f), the tiles
 which are decorated in this Figure must repeat with period $4\ZZ^2$.
If $C$ is the undecorated square underlying $S_0$, viewed as a
 subset of $\RR^2$, we must have, as a consequence of the self-similarity
 involved in the definition of ${\cal T}_0$, that
\be
  (P_k\cap T^i C) + 2^i \cdot 4\ZZ^2 \subseteq P_k
\ee
 for all $i\in\NN$, $k\in\{ 0,1,2,3\}$.
Therefore, if we set $P_{k,i}:= P_k\cap T^i C$, we get
\be\label{chairdecom}
 P_k \; = \; \bigcup_{i\in\NN}\;
             \bigcup_{t\in P_{k,i}}(t+2^i \cdot 4\ZZ^2) \;,
\ee
 which is the desired decomposition of $P_k$.

Using the substitution rule, the finite sets $P_{k,i}:= P_k\cap T^i C$
 can actually be calculated by recursion:
\be
 P_{0,0}\; = \;\{ 0\}\, , \zr{0.5} P_{k,0}\; = \;\emptyset\, , \; k\in\{ 1,2,3\},
\ee
\be
P_{k,i+1}\; = \;\bigcup_{l=0}^3 \, T^i M_l T^{-i} (P_{(k-n_l)_4,i})
\ee
 for the integers $n_0:=0$, $n_1:=1$, $n_2:=2$, $n_3:=1$ and
 the affine transformations $M_l$ given by
$M_0 x:=x$, 
$M_1 x:= Rx+\vec{\bm{e}}_1$,
$M_2 x:= R^2x+\vec{\bm{e}}_1+\vec{\bm{e}}_2$,
$M_3 x:= Rx+\vec{\bm{e}}_2$.

In much the same way as in the one-dimensional example, the decomposition
 (\ref{chairdecom}) leads to a description of $P_k$ as model sets.
The lattice $\ZZ^2$ is embedded in $G$ in the canonical fashion.
In the embedding space  $\RR^2\times G$ we choose the lattice 
 $L:= \{ (\vec{n}, \vec{n}) | \vec{n}\in  \ZZ^2\}$.
For each $i \in \NN$, the closure with respect to the 2-adic topology
 of the sublattice $2^i \cdot 4\ZZ^2$ is an open and
 compact subgroup of $G$, actually equal to $2^i \cdot 4 G$.
Replacing $\ZZ^2$ by $G$ in (\ref{chairdecom}) and taking the 2-adic 
closure gives the description of windows $\Omega_k\subseteq G$ in $G$.
It is easily seen that $P_k$ is the model set using $G$ as internal space,
 $L$ as lattice and $\Omega_k$ as window.

{}Finally, we show that the boundary of each open set $\Omega_k$ has
Haar measure 0. Let $\mu$ be the Haar measure on $G$ and assume it is
normalized to $\mu(G)=1$. Then the measure of any coset
$t+2^i\cdot 4\cdot G$ is $1/[G:(2^i\cdot 4 \cdot G)]$.
On the other hand, the proportion of points of $\ZZ^2$ lying in the
coset $t+2^i\cdot 4 \ZZ^2$ is $1/[\ZZ^2 : (2^i \cdot 4 \ZZ^2)]$ 
which is exactly the same number.
Then the proportion of points of $\ZZ^2$ occupied by the cosets
belonging to $P_k$ is exactly the same as $\mu(\Omega_k)$.
Since $\bigcup_{k=0}^{3} P_k = \ZZ^2$ we obtain 
$\sum_{k=0}^{3} \mu(\Omega_k) = 1$. From 
$\overline{\Omega}_0 \cap (\Omega_1 \cup \Omega_2 \cup \Omega_3) = \emptyset$
we have 
$1 \geq \mu(\overline{\Omega}_0) + \sum_{k=1}^{3} \mu(\Omega_k)
   \geq \sum_{k=0}^{3} \mu(\Omega_k) = 1$
and so $\mu(\overline{\Omega}_0) = \mu(\Omega_0)$.
Similarly, one gets $\mu(\partial\Omega_k)=0$ for $k=0,1,2,3$ as required.

\section{A limit-quasiperiodic example} \label{lqss}

The substitution matrix of the primitive two letter substitution system
\be\label{limitsub}
 a \rightarrow aab \qquad b \rightarrow abab
\ee
 has the Perron-Frobenius eigenvalue $\lambda:=2+\sqrt{2}$ which is a 
Pisot-Vijayaraghavan number but not a unit.
Therefore, any possible description as a model set of the
 resulting substitution sequence will have to use more complicated
 groups than $\RR^n$ as embedding space.

A geometric representation of the substitution system is obtained
 by replacing symbols $a$ and $b$ by intervals of length $\ell(a)=1$ and
 $\ell(b)=\sqrt{2}$.
If we denote the sets of left endpoints of the $a$ and $b$ intervals
 by $\Lambda_a$ and $\Lambda_b$, the substitution
 rule leads to the following system of recursion relations:
\begin{eqnarray}\label{qprek}
\Lambda_a & = & (\lambda \Lambda_a) \cup
                (\ell(a)+\lambda \Lambda_a) \cup
                (\lambda \Lambda_b) \cup
                (\ell(a)+\ell(b)+\lambda \Lambda_b) \label{rel1} \\
\Lambda_b & = & (2\ell(a)+\lambda \Lambda_a) \cup
                (\ell(a)+\lambda \Lambda_b) \cup
                (2\ell(a)+\ell(b)+\lambda \Lambda_b) \label{qprek2}
\end{eqnarray}
 where the right hand sides represent disjoint unions.
Both $\Lambda_a$ and $\Lambda_b$ are
 subsets of the group $\ZZ[\sqrt{2}]$ which can be mapped onto
 $\ZZ^2\subseteq\RR^2$ by sending $\ell(a)$ to $\vec{\bm{e}}_1$ and
 $\ell(b)$ to $\vec{\bm{e}}_2$.
The transformation $t\mapsto\lambda t$ in $\ZZ[\sqrt{2}]$ induces
 the linear transformation 
 $\phi: \vec{\bm{e}}_1 \mapsto  2 \vec{\bm{e}}_1+\vec{\bm{e}}_2,
 \mbox{\hspace{2mm}} \vec{\bm{e}}_2\mapsto  2 \vec{\bm{e}}_1+2\vec{\bm{e}}_2$ 
 of $\RR^2$.
Notice that $\phi(\ZZ^2)\subset\ZZ^2$, but $\phi(\ZZ^2)\neq\ZZ^2$.

The transformation $\phi$ has the two eigenvalues $2\pm\sqrt{2}$;
 we may identify the ``physical'' space $\RR$ with the subspace $V$
 of $\RR^2$ corresponding to the eigenvalue $2+\sqrt{2}$.
Then, all points $\Lambda_a$ and $\Lambda_b$ of a substitution
 sequence according to (\ref{limitsub}) are images of uniquely
 determined points of $\ZZ^2$ under the projection onto the
 physical space $V$ along the second invariant subspace of $\phi$.
(Note that, as the transformation matrix of $\phi$ is not normal,
 this projection is not orthogonal with respect to the
 canonical metric of $\RR^2$; see Figure 2.)
Because the second eigenvalue is smaller than 1, the preimages
 must lie in a bounded strip parallel to $V$, and it is easily
 calculated that, for the sequence generated from $ba$, where the
 middle vertex is the origin, the preimages lie in the strip
 $V+\{ t(0,-1-\sqrt{2})\, |\, 0\leq t \leq 1 \}$
 (see Figure 2).

The problem is that not all points of $\ZZ^2$ which are in the
 strip are preimages of points in the sequence, therefore
 the embedding so far does not exhibit the sequence as a
 model set.
However, there is an open substrip whose model set is
 a subset of the sequence.
This can be seen by observing that the preimages of the sequence
 can be connected by a path which only passes along horizontal
 and vertical bonds in the square lattice (cf.\ Figure 2).
If one would omit any point of $\ZZ^2$ in the strip 
\be \label{strip}
  V+\{ (0, -\sqrt{2}/2 ) + t (0,-1-\sqrt{2}/2)\, |\, 0\leq t \leq 1 \} \, ,
\ee
 then no such connected path would be possible any more.
This observation is the analogue of finding periodic subsets
 in the limit periodic examples; all further steps are
 more or less completely determined.

We extend $\RR^2$ by the inverse limit $G :=\overleftarrow{\,\ZZ}{}^2(\phi)$
 (see Eq.~(\ref{profinite})) and embed $\ZZ^2$ in $\RR^2 \times G$ in the 
 canonical fashion as the lattice $L$.
The homomorphism $\phi$ can be uniquely extended to $\RR^2 \times G$.
From the above considerations, if we take as internal group $G'$ the
 product of the second eigenspace of $\phi$ with $G$, we find an open
 window in $G'$ such that the corresponding model set is a subset
 of the substitution sequence $\Lambda:=\Lambda_a\cup\Lambda_b$.
The recursion relations (\ref{qprek}) and (\ref{qprek2})
 can be transfered to the internal group $G'$.
This gives an iterated function system for two windows $\Omega_a$ and
 $\Omega_b$ related to type $a$ and $b$ vertices.
This system has a unique pair of compact sets 
 $(\Omega_a,\Omega_b)$ as its attractor.

Obviously, $\Lambda\subseteq\Lambda(\Omega)$ for
 $\Omega:=\Omega_a\cup\Omega_b$.
Due to the recursion relation (\ref{rel1}), $\lambda\Lambda$ is
 a subset of $\Lambda_a$.
From the observation (\ref{strip}) we can find an open subset of $G$
inside $\Omega_a$. 
Then, using the recursion relation both for $\Lambda_{a,b}$ and $\Omega_{a,b}$
 we can find an open set $U\subset \Omega$
 which has $\Omega$ as its closure such that $\Lambda(U)\subseteq \Lambda
 \subseteq\Lambda(\Omega)$.
Since $\Omega\backslash U$ has no interior, $\Lambda(U)$ and $\Lambda(\Omega)$
 differ only on a set of points that is {\em not} relatively dense, i.e.\
 on a set (in $\RR$) that has gaps of arbitrary length.

By the argument at the end of the Introduction, we can find a $c\in G'$ so that
 $\Lambda(c+\Omega)$ is regular, and hence repetitive, as is the original
 substitution sequence $\Lambda$.
Now let us show that $\Lambda$ and $\Lambda(c+\Omega)$ are locally isomorphic,
 thus establishing $\Lambda$ to be a model set as defined in the Introduction.

From the above argument, we know that there are arbitrarily long intervals 
 where $\Lambda(U)$, $\Lambda$ and $\Lambda(\Omega)$ coincide.
Let us select such an interval of length $R$. Then, we also know that
 there exists a relatively dense set of translations $t$ such that
 $t + \Lambda(c+\Omega)$ coincides both with $\Lambda(U)$ and
 $\Lambda(\Omega)$ on that interval, too.
Since $\Lambda$ is repetitive, this establishes that $\Lambda$ and
 $\Lambda(c+\Omega)$ are locally isomorphic.

This finally reveals the points $\Lambda_a$ and $\Lambda_b$ as model sets
 based on the mixed internal space $\RR \times \overleftarrow{\,\ZZ}{}^2(\phi)$.

\begin{figure}
\vspace*{5mm}
\centerline{\epsfysize=80mm \epsfbox{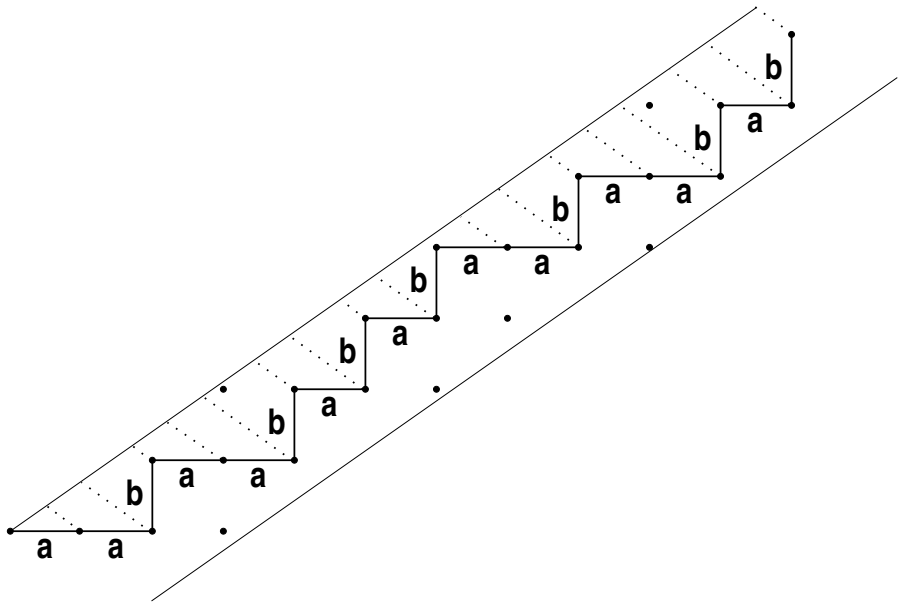}}
\caption{The limit quasiperiodic example.}\label{fig2}
\end{figure}

\section{Diffraction}

The diffraction of a generalized model set can be calculated
 in much the same way as for model sets in the conventional
 framework \cite{forthcoming}.
Given a model set $\Lambda$ in $\RR^d$, one can show that its characteristic
 Dirac comb, i.e.\ the measure
\begin{equation}
     \omega \; = \;  \omega_{\Lambda} 
           \; := \; \sum_{t\in\Lambda}\delta_t \, ,
\end{equation}
has a unique autocorrelation,
\begin{eqnarray}
 \gamma \; = \; \gamma^{}_{\omega} 
        & := &  \lim_{r\rightarrow\infty} 
               \frac{1}{{\rm vol}(B_r(0))} (\omega_{\Lambda_r} *
                              \tilde{\omega}_{\Lambda_r})  \nonumber \\
      & = & \lim_{r\rightarrow\infty} \frac{1}{{\rm vol}(B_r(0))}
               \sum_{s,t\in\Lambda_r}\delta_{t-s}  \, ,
\end{eqnarray}
where $B_r(0)$ is the ball of radius $r$ around $0$,
$\Lambda_r := \Lambda \cap B_r(0)$ and $\tilde{\omega}$ denotes the
measure defined by $(\tilde{\omega},\phi)=\overline{(\omega,\tilde{\phi})}$
with $\tilde{\phi}(x):=\overline{\phi(-x)}$. 

A good theory of diffraction exists for model sets under the additional
assumption
\begin{itemize}
\item[\bf W2] The boundary of $\Omega$ has measure 0 (measure
               being the Haar measure of $G$).
\end{itemize}
If internal space is Euclidean, this is tantamount to saying that the
window is a Riemann measurable set.

If one interprets the measure $\omega$ as a set of point scatterers at the
 sites of $\Lambda$, then the corresponding diffraction pattern
 is the Fourier transform $\hat{\gamma}$ of the autocorrelation $\gamma$.
This Fourier transform $\hat{\gamma}$ is itself a positive measure and has,
 for general model sets $\Lambda$ with property {\bf W2}, only a point component, 
 i.e., can be written in the form
\begin{equation}\label{pointpart}
  \hat{\gamma} \; = \; \sum_{k\in F} C(k)\delta_k
\end{equation}
 with non-negative coefficients $C(k)$.
The set $F$ in (\ref{pointpart}) is the projection into $\RR^d$ of the
 dual lattice of $L$ in  the dual of $\RR^d \times G$.

Without  going into the detailed calculation, let us give the
 result in the case of the limit-periodic substitution sequence (\ref{substRule})
 for which we can easily verify condition {\bf W2} -- indeed, the boundary of
 the window is a finite point set.
We denote the right-hand end points of the intervals $a$, $b$ and
 $c$ by $\Lambda_{a}$, $\Lambda_{b}$ and $\Lambda_{c}$, 
 and consider them as the positions of point scatterers of strengths 
 $h_a$, $h_b$ and $h_c$, respectively.
The Fourier transform of the autocorrelation $\gamma^{}_{\omega}$ of the measure
\begin{equation}\label{measure}
   \omega  \; = \; 
         h_a \sum_{t\in \Lambda_a} \delta_t+
         h_b \sum_{t\in \Lambda_b} \delta_t+
         h_c \sum_{t\in \Lambda_c} \delta_t
\end{equation}
is then given by
\begin{equation}\label{diffraction}
 \hat{\gamma} \; = \; 
              \sum_{k\in F} |h_a A_a(k) + h_b A_b(k) + h_c A_c(k)|^2 \delta_k
\end{equation}
 with the ``amplitudes''
\begin{eqnarray}\label{amplitudes}
A_a(k)&=&
 \frac{1}{3^n} e^{\pi i m/3^n}\left(e^{-\pi i m / 3} + \frac{(-1)^m}{2}
\right)\, ,  \nonumber
 \\ 
A_b(k)&=&
 \frac{1}{3^n} e^{-\pi i m/3^{n-1}}\left(e^{-\pi i m / 3} + \frac{(-1)^m}{2}
\right)\, ,
 \\ \nonumber
A_b(k)&=&
 \frac{1}{3^n} e^{-\pi i m/3^{n-1}}\left(e^{\pi i m / 3} + \frac{(-1)^m}{2}
\right) \, .
 \\ \nonumber
\end{eqnarray}
The sum in (\ref{diffraction}) runs over the Fourier module $F$, 
\begin{equation}\label{ftmodule}
  F \; := \; \left\{ k = \frac{m}{3^n} \; \mid \;
         (n=2, m\in\ZZ) \mbox{ or }
         (n\geq 3, m\nqv 0 \;\mbox{ mod}(3))\, \right\} \, ,
\end{equation}
namely the set of all rational numbers $k$ whose denominators 
are, at worst, powers of $3$. Each such number $k$ is uniquely expressible 
in the form indicated in (\ref{ftmodule}).
It is this one-to-one parameterization that appears in (\ref{amplitudes}).

It is easy to see that $F$ is indeed the
 projection into $\RR$  of the dual of $\ZZ$ in the dual of
 $\RR\times\widehat{\ZZ_3}$.

In the case of the chair tiling, we already established {\bf W2}, and the
diffraction can be calculated along similar lines to the previous example.
The limit-quasiperiodic substitution system of Section~\ref{lqss} is a lot
more complicated, and we have not even been able to verify {\bf W2} so far.

\section{Comments}

The formalism of model sets has been shown to encompass
situations not hitherto considered within 
its scope by using internal spaces with non-Euclidean topologies.
The situations in which such topologies occur are signalled
by the presence of chains of decreasing sublattices of ever increasing
scale. 

Let us point out a few more examples.
The period doubling substitution rule $a\rightarrow ba$,
$b\rightarrow aa$, which is known to have pure point spectrum
from Dekking's criterion \cite{dekking}, gives rise to a $2$-adic model set.
One of the oldest aperiodic tilings is the Robinson tiling
\cite{Robinson}, which is based on a set of six decorated squares. This
tiling has an interpretation in terms lattices of overlapping squares
(see \cite{GS} for a picture) which clearly shows its $2$-adic
nature (something already realized by Robinson). In fact, the centers
of the tiles of each type form a $2$-adic model set. In the course of
working out these examples, we discovered that the chair tilings
and the Robinson square tilings are actually closely related. Suitably
decorated, the chair tiling can be transformed into a Robinson tiling
and in the reverse direction, suitably undecorated, the Robinson
tiling can be transformed into a chair tiling. 

The sphinx tiling as well as the new hexagonal tiling of Penrose 
\cite{Penrose} undoubtedly also admit $2$-adic interpretations. 

The substitution tilings described above have so far
 been considered as belonging to the classes of aperiodic
 tilings called limit-periodic and limit-quasiperiodic tilings,
 compare \cite{GK}. 
Potential limit-(quasi-)periodic tilings can be recognized by displaying
 an inflation/deflation symmetry in the sense of \cite{BS} with
 an inflation multiplier that is an algebraic integer larger than 1,
 but not a unit.
Not all of them will display a pure point diffraction spectrum, as can
 be seen from the Thue-Morse chain (defined by $a\rightarrow ab$,
 $b\rightarrow ba$) or a variant of our system (\ref{substRule}) (defined by
 $a\rightarrow ab$, $b\rightarrow abc$, $c\rightarrow ccab$).
These cases cannot be model sets.
Our analysis shows nevertheless that a unified description
 of at least some of the cases with pure point spectrum is possible if 
 one slightly generalizes the class of internal spaces which are admitted.

This generalization turns out to be a very natural one.
Many properties of conventional model sets can be proved
 to hold also in the more general case.
Among these are the uniform densities of general model
 sets (see \cite{Martin2}) and, if also {\bf W2} is fulfilled,
 the pure point character
 of the diffraction spectrum (see \cite{forthcoming}).
It would be nice to find an exhaustive criterion for those cases
 with pure point diffraction spectrum.

\ack

The authors thank L. Danzer for
his interest and helpful discussions on the chair tiling. 
RVM is grateful to the Natural Sciences and
Engineering Council of Canada for the continuing support
of his research. MS thanks the Pacific Institute of Mathematical
Sciences for support of his research in Canada.


\section*{References}

\begin{thebibliography}{99}
\small

\bibitem{BS}
M.~Baake and M.~Schlottmann,
 ``Geometric Aspects of Tilings and Equivalence Concepts'',
 in: {\em Proc.\ of the 5th Int.\ Conf.\ on Quasicrystals},
 eds.\ C.~Janot and R.~Mossery,
 World Scientific, Singapore (1995), pp.\ 15--21.

\bibitem{Bourbaki}
N.~Bourbaki, {\em Topology}, Vol.~1, Addison-Wesley, Reading (1966).

\bibitem{deBr}
N.~G.~de Bruijn, 
 ``Algebraic theory of Penrose's non-periodic tilings of the
 plane'', part I: {\em Math.\ Proc.} {\bf A84} (1981) 39--52, and part II:
 {\em Math.\ Proc.} {\bf A84} (1981) 53--66.

\bibitem{dekking}
F.~M.~Dekking,
``The spectrum of dynamical systems arising from substitutions of constant
 length'',
{\em Z.\ Wahrscheinlichkeitstheorie} {\bf 41} (1978) 221--239.

\bibitem{GK}
F.~G\"ahler and R.~Klitzing,
``The diffraction pattern of self-similar tilings'', in:
{\em The Mathematics of Long-Range Aperiodic Order},
 ed.\ R.~V.~Moody,
 NATO ASI Series C 489, Kluwer, Dordrecht (1997),
 pp.\ 141--74.

\bibitem{GS}
B.~Gr\"unbaum and G.~C.~Shephard, {\em Tilings and Patterns},
Freeman, New York (1987).

\bibitem{HR}
E.~Hewitt and K.~A.~Ross, {\em Abstract Harmonic Analysis},
Vol.~1, 2nd ed., Springer, New York (1979).

\bibitem{bert}
A.~Hof,
``Diffraction by aperiodic structures'', in:
 {\em The Mathematics of Long-Range Aperiodic Order},
 ed.\ R.~V.~Moody,
 NATO ASI Series C 489, Kluwer, Dordrecht (1997), pp.\ 239--68.

\bibitem{Katz}
A.~Katz and M.~Duneau,
``Quasiperiodic patterns and icosahedral symmetry'',
{\em J.\ Physique} {\bf 47} (1986) 181--96.

\bibitem{Kramer}
P.~Kramer,
``Non-periodic central space filling with icosahedral symmetry using
copies of seven elementary cells'',
{\em Acta Cryst.} {\bf A38} (1982) 257--64.

\bibitem{KN}
P.~Kramer and R.~Neri,
``On periodic and non-periodic space fillings of $\EE^m$ obtained by projection'',
{\em Acta Cryst.} {\bf A40} (1984) 580--7, and
{\em Acta Cryst.} {\bf A41} (1985) 619 (Erratum).

\bibitem{McM}
P.~McMullen,
 ``Duality, sections and projections of certain Euclidean tilings'',
 {\em Geom.\ Dedicata} {\bf 49} (1994) 183--202.

\bibitem{YM1}
Y.~Meyer, {\em Algebraic Numbers and Harmonic Analysis},
North-Holland, Amsterdam (1972).

\bibitem{RVM}
R.~V.~Moody, ``Meyer sets and their duals'', in:
{\em The Mathematics of Long-Range Aperiodic Order},
ed.\ R.~V.~Moody, NATO ASI Series C 489,
Kluwer, Dordrecht (1997), pp.\ 403--41.

\bibitem{Neukirch}
J.~Neukirch, ``The $p$-adic numbers'', in: {\em Numbers},
eds.\ H.-D.\ Ebbinghaus et al., Springer, New York (1990), pp.\ 155--178.

\bibitem{Penrose}
R.~Penrose, ``Remarks on tiling: Details of a
$(1 \!+\! \epsilon \!+\! \epsilon^2)$-aperiodic set'', in:
{\em The Mathematics of Long-Range Aperiodic Order},
ed.\ R.~V.~Moody, NATO ASI Series C 489, Kluwer,
Dordrecht (1997), pp.\ 467--97.

\bibitem{queff}
M.~Queffelec,
``Spectral study of automatic and substitutive sequences'',
 in: {\em Beyond Quasicrystals}, ed.\ F.~Axel and D.~Gratias, Springer, Berlin
 (1994), pp.\ 369--414.

\bibitem{Robinson}
R.~M.~Robinson, ``Undecidability and nonperiodicity of tilings
of the plane'', {\em Inv.\ Math.} {\bf 44} (1971) 177--209.


\bibitem{Martin2}
M.~Schlottmann, ``Cut-and-project sets in locally compact
Abelian {groups}'', in: 
{\em Quasi\-crystals and Discrete Geometry}, ed.\ J.~Patera,
Fields Institute Monographs, Vol.~10, AMS, Rhode Island (1998), pp.\ 247--64.

\bibitem{forthcoming}
M.~Schlottmann, ``Generalized model sets and dynamical systems'',
to appear in: {\em Directions in Mathematical Quasicrystals},
eds.\ M.~Baake and R.~V.~Moody,
CRM monograph series, AMS, Rhode Island (1998), in preparation.

\bibitem{Serre}
J.-P.~Serre, {\em A Course in Arithmetic}, Springer, New York (1973).

\bibitem{boris}
B.~Solomyak, ``Dynamics of self-similar tilings'',
 {\em Ergod.\ Th.\ \& Dynam.\ Syst.} {\bf 17} (1997) 695--738.

\end{thebibliography}
\end{document}